\documentclass[superscriptaddress,prb,twocolumn,showpacs,preprintnumbers,amsmath,amssymb,twoside]{revtex4-1}

\usepackage{graphicx}		
\usepackage{bm}
\usepackage{natbib}
\usepackage{color}

\definecolor{SBW}{rgb}{1,0,0}

\begin{document}

\newcommand{\etal}{{\it et al.}}
\newcommand{\tco}{T$_{\mbox{\scriptsize CO}}$}
\newcommand{\tso}{T$_{\mbox{\scriptsize SO}}$}
\newcommand{\tc}{T$_{\mbox{\scriptsize C}}$}
\newcommand{\tilt}{I$_{\mbox{\scriptsize 300}}$}
\newcommand{\YMO}{YMn$_{\mbox{\scriptsize 2}}$O$_{\mbox{\scriptsize 5}}$}
\newcommand{\TMO}{TbMn$_{\mbox{\scriptsize2}}$O$_{\mbox{\scriptsize 5}}$}
\newcommand{\LBCOF}{La$_{1.875}$Ba$_{0.125}$CuO$_4$}
\newcommand{\NdLSCOF}{La$_{1.48}$Nd$_{0.4}$Sr$_{0.12}$CuO$_4$}
\newcommand{\LBCO}{LBCO}
\newcommand{\NdLSCO}{LNSCO}


\title{A Comparison of Stripe Modulations in \LBCOF\ and \NdLSCOF}

\author{S.B. Wilkins}
\email{swilkins@bnl.gov}
\author{M. P. M. Dean}
\affiliation{Condensed Matter Physics and Materials Science
  Department, Brookhaven National Laboratory, Upton, New York, 11973
  USA}
\author{J\"org Fink}
\affiliation{Leibniz-Institute for Solid State and Materials Research
  Dresden, P. O. Box 270116, D-01171 Dresden, Germany}
\affiliation{Helmholtz-Zentrum Berlin f\"ur Materialien und Energie,
  Albert-Einstein-Stra\ss e 15, D-12489 Berlin, Germany}
\author{Markus H\"ucker}
\affiliation{Condensed Matter Physics and Materials Science
  Department, Brookhaven National Laboratory, Upton, New York, 11973
  USA}
\author{J.~Geck}
\affiliation{Leibniz-Institute for Solid State and Materials Research
  Dresden, P. O. Box 270116, D-01171 Dresden, Germany}
\author{V. Soltwisch}
\author{E. Schierle}
\author{E. Weschke}
\affiliation{Helmholtz-Zentrum Berlin f\"ur Materialien und Energie,
  Albert-Einstein-Stra\ss e 15, D-12489 Berlin, Germany}
\author{G. Gu}
\affiliation{Condensed Matter Physics and Materials Science
  Department, Brookhaven National Laboratory, Upton, New York, 11973
  USA}
\author{S. Uchida}
\affiliation{Department of Physics, University of Tokyo, Tokyo 113-8656, Japan}
\author{N. Ichikawa}
\affiliation{Institute for Chemical Research, Kyoto University, Gokasho, Uji 610-0011, Japan}
\author{J.M. Tranquada}
\author{J.P. Hill}
\affiliation{Condensed Matter Physics and Materials Science
  Department, Brookhaven National Laboratory, Upton, New York, 11973 USA}

\date{\today}

\begin{abstract}
We report combined soft and hard x-ray scattering studies of the
electronic and lattice modulations associated with stripe order in  \LBCOF\ and
\NdLSCOF.  We find that the amplitude of both the electronic modulation of the hole density and the strain modulation of the lattice is significantly larger in  \LBCOF\ than in \NdLSCOF\ and is also better correlated. The in-plane correlation lengths are isotropic in each case; for \LBCOF,
$\xi^{hole}=255\pm 5$~\AA\ whereas for \NdLSCOF,  $\xi^{hole}=111\pm 7$~\AA.
We find that the modulations are temperature independent in \LBCOF\ in
the low temperature tetragonal phase. In contrast, in \NdLSCOF, the amplitude grows smoothly from zero, beginning 13~K below the LTT phase transition.
We speculate that the reduced average tilt angle in \LBCOF\ results in reduced charge localization and incoherent pinning, leading to the longer correlation length and enhanced periodic modulation amplitude.

\end{abstract}


\maketitle

One dimensional modulations of the lattice and of the charge and spin density appear to be both ubiquitous in the cuprates and intimately tied up with the physics of these materials. For example, superconducting phase coherence is suppressed in systems in which charge and spin stripes are static\cite{PhysRevLett.78.338,PhysRevB.38.4596, PhysRevB.70.104517} while dynamic stripes may promote pairing\cite{Kivelson:1998ir,PhysRevB.56.6120}. Further, stripe order coexists with 2D superconducting correlations above the bulk superconducting phase\cite{PhysRevLett.99.067001,PhysRevB.78.174529} and indeed it has been argued that broken rotational symmetry may be a defining feature of the pseudogap phase - so-called  ``nematic'' electronic order, a viewpoint for which there is some evidence\cite{Lawler:1969ei,Daou:2031bo} . Understanding these modulations is therefore a prerequisite to gaining a detailed knowledge of the physics of the cuprates.

In their full incarnation, these density waves - ``stripes'' -  consist of three concomitant modulations; of the spin density, of the charge density and of the lattice itself. In a few cases, namely La$_{2-x}$Ba$_x$CuO$_4$\cite{PhysRevB.70.104517,Abbamonte:2005bl,Kim:2008jn,PhysRevB.78.174529,Hucker:2011je} and (La,R)$_{2-x}$Sr$_x$CuO$_4$, where R=Nd\cite{Tranquada:1995kx}, or Eu \cite{PhysRevB.79.100502,PhysRevB.83.092503}
, these modulations are stabilized by the symmetry of the low temperature crystal structure to form static structures. It is here, where the respective order parameters are well developed, that one can best characterize them. The basics of the static structures are known: they are comprised of hole-rich anti-phase antiferromagnetic domain walls running along the Cu-O bond direction. In real space the spin density then has twice the wavelength of the charge and lattice modulations, and in a scattering experiment, these give rise to satellites at ($0.5\pm \delta$,$0.5$,0) and ($\pm 2\delta$,0,0.5), and symmetry related positions, respectively. However, basic questions such as the relative amplitudes in the different compounds and the coupling between these modulations remain unanswered.

Here we report a combined soft and hard x-ray scattering study of the
low temperature stripe order in \LBCOF\ (\LBCO) and \NdLSCOF\ (\NdLSCO). In particular, soft
x-ray scattering at the pre-edge of the O K-edge (529.3~eV) probes the
spatial variation of the oxygen $2p$ partial density of states at the Fermi level, which is related to the mobile charge
carrier
density\cite{PhysRevLett.66.104,PhysRevB.42.8768,Abbamonte:2005bl},
while hard x-ray (11.3~keV)
scattering probes the associated lattice distortion. We find that the
stripe correlation lengths are isotropic in the Cu-O plane in both
systems, but \LBCO\ is a factor of two better correlated. Further, by
careful normalization to structural Bragg peaks we are able to
quantify the relative size of the electronic modulation and the
lattice distortion in these two cases. We find that the amplitude of the
electronic modulation is $\sim 10$ times larger in \LBCO\ than in
\NdLSCO, while the lattice distortion is only $\sim 4$ times
larger. We further find that the modulations are temperature
independent in \LBCO\ once the LTT phase is entered, suggesting that perhaps
the intrinsic ordering temperature is significantly higher. In
contrast, in \NdLSCO, the amplitude grows smoothly from zero,
beginning 13 K below the LTT phase transition.

Samples of both \LBCO\ and \NdLSCO\ were grown using the
floating zone method, at Brookhaven
National Laboratory, USA and the University of Tokyo,
Japan respectively. Both samples proved to be of high crystal quality, with
mosaic spreads of $\le 0.02^\circ$ measured at 11.3~keV. Throughout this paper both systems are indexed with the high temperature tetragonal (HTT) ({\it I4/mmm}) unit cell.

Soft x-ray diffraction experiments were carried out on the X1A2 beamline at the NSLS, Brookhaven National Laboratory, USA using a six-circle in-vacuum diffractometer. The samples were cleaved \emph{ex-situ} to reveal surfaces with a $[001]_\mathrm{HTT}$ surface normal and mounted such that the $[001]_\mathrm{HTT}$ and $[100]_\mathrm{HTT}$ directions lay in the scattering plane. Experiments were performed in a vertical scattering geometry with $\sigma$-incident x-rays, i.e. the electric field of the incident x-ray photons was always along the HTT $b$-axis, within the CuO$_2$ planes. The samples were cooled in a He flow cryostat to a base temperature of 12~K. Scattered x-rays were detected with an in-vacuum CCD. During the measurements the beam footprint was smaller than the sample. Data were collected by performing a  $\theta-2\theta$ scan, collecting a single CCD image at each datum. Each CCD image was then converted to reciprocal space coordinates and  binned onto a regular grid \cite{ccd:unpublished}. Part of the soft x-ray experiments were performed at BESSY with a diffractometer described in Ref.~\onlinecite{PhysRevB.79.100502}.

Hard x-ray diffraction experiments on the same samples were conducted on beamline X22C at the NSLS. The incident photon energy was 11.3~keV  and scattered x-rays were detected by means of a point-detector, with the resolution defined by slits. Experiments were conducted in a vertical scattering geometry, with the $[001]_\mathrm{HTT}$ and $[100]_\mathrm{HTT}$ directions lying in the scattering plane. For both \LBCO\ and \NdLSCO\ the instrumental resolution was identical.

\begin{figure}[t!]
  \includegraphics[width=0.8\columnwidth]{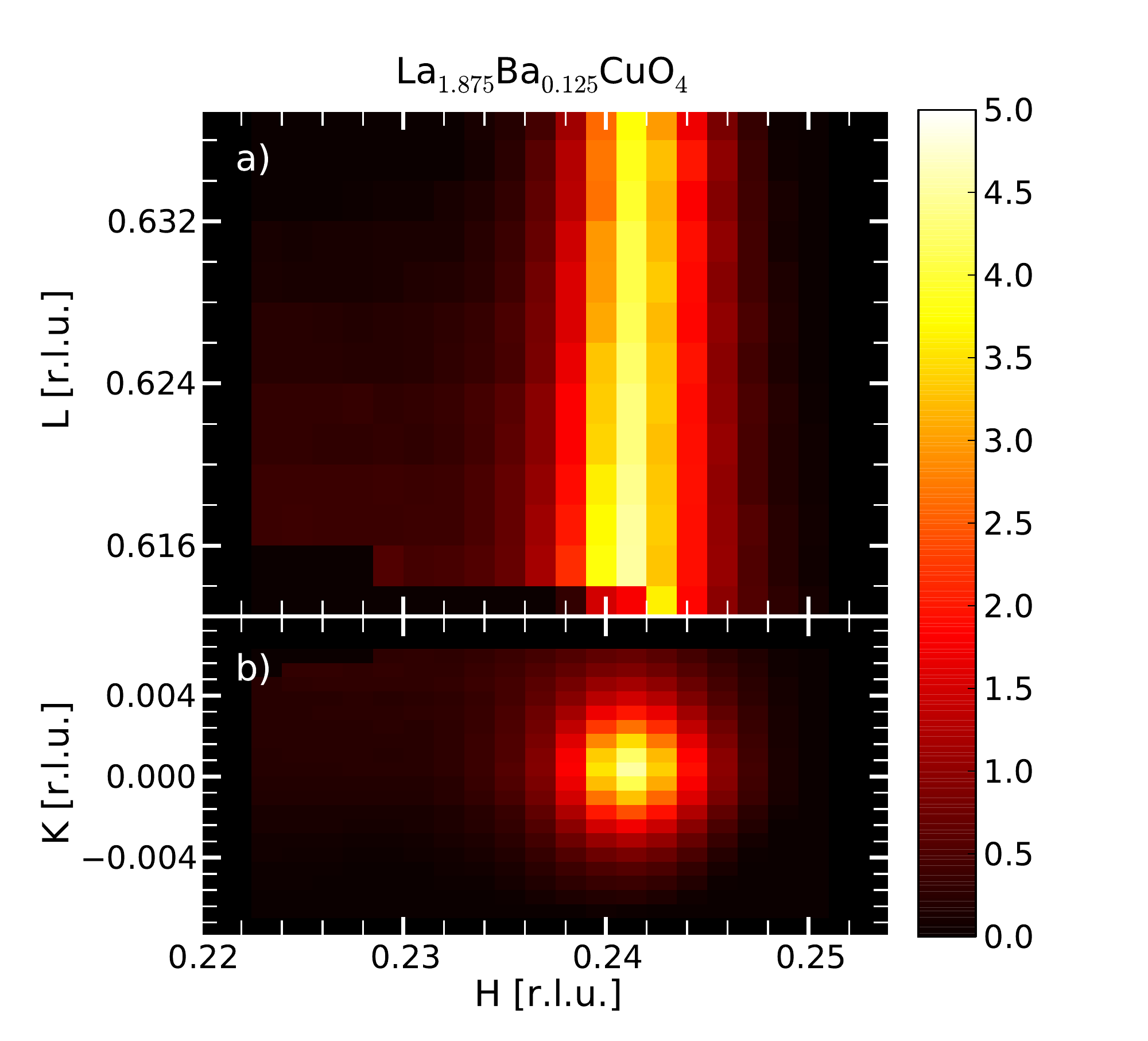}
  \caption{Soft x-ray resonant scattering intensity in the vicinity of the $(0.24, 0, L)_\mathrm{HTT}$ stripe order peak for \LBCO\ in the a) $(H,0,L) _\mathrm{HTT}$ plane and b) $(H,K,0.624) _\mathrm{HTT}$ plane in reciprocal space. Data were taken at T=15 K, and E$_i$=529.3 eV. At this energy, the scattered intensity reflects the modulation of the oxygen density of states in real space. The intensity map reveals that the stripe correlations are isotropic in the \emph{a-b} plane and are two dimensional.}
  \label{fig:lbco-hkl}
\end{figure}

We first discuss the characterization of the electronic modulation,
beginning with \LBCO. In Fig.~\ref{fig:lbco-hkl} we show cuts through
reciprocal space in the HK- (bottom) and HL- planes (top) for
\LBCO. These data were taken with E$_i$=529.3 eV, an energy
corresponding to the Mobile Carrier Peak (MCP) in the oxygen K-edge
absorption\cite{PhysRevLett.66.104}. At this energy the scattering
from the charge stripes is maximized\cite{Abbamonte:2005bl}. We
extract a correlation length from these data by taking line cuts
through the three dimensional grid and fitting to Lorentzian-squared
lineshapes. We find that the stripe correlations within the $a$-$b$
plane are isotropic with $\xi_H = 255\pm5$~\AA, where
$\xi=\frac{1}{HWHM}$.  Measurements at the Cu L$_3$
edge revealed a similar correlation length. We note that because of the superior
resolution of the present set up, this represents the highest precision measurement
to date of this correlation length. It is consistent with previous
work which found $\xi \approx~200$~\AA\ at this
energy\cite{Abbamonte:2005bl}.  As has been previously observed, the
stripes are weakly correlated in the $c$-axis direction, giving rise
to the uniform streak of intensity along $L$ in
Fig.~\ref{fig:lbco-hkl}.

\begin{figure}
  \includegraphics[width=0.8\columnwidth]{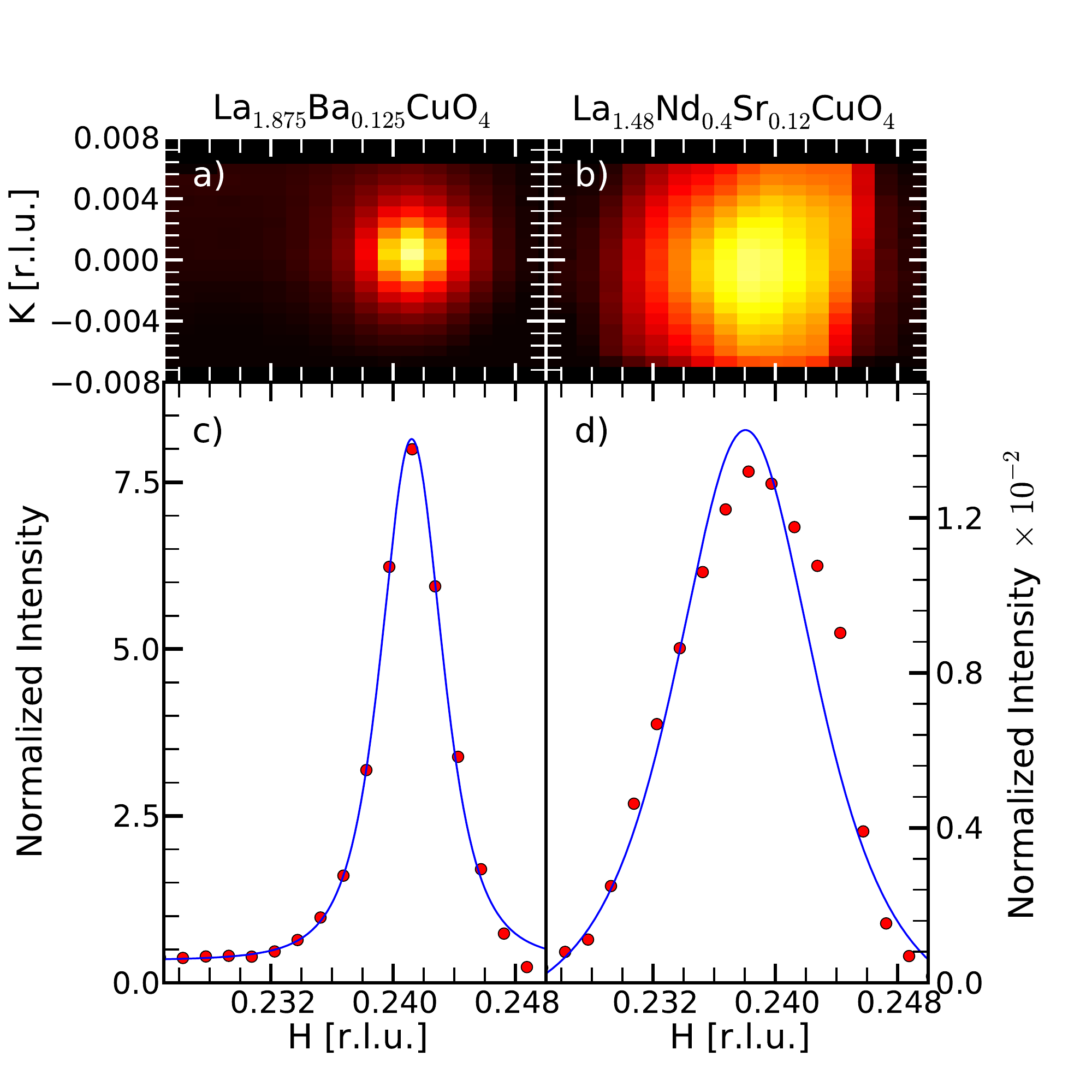}
  \caption{Comparison of scattering from stripe modulations in \LBCO\ and \NdLSCO. Data were taken at T=15 K, and E$_i$=529.3 eV. a) and c) $(H,K,0.624)_\mathrm{HTT}$ plane and $(H,0,0.624)_\mathrm{HTT}$ scan (circles) for \LBCO. b) and d) same for \NdLSCO. In c) and d) the solid line is the result of a fit to a Lorentzian squared lineshape. Scattered intensities are normalized to the respective $(002) _\mathrm{HTT}$ Bragg peaks and  therefore may be compared directly to each other.}
  \label{fig:comp-soft}
\end{figure}

Figure~\ref{fig:comp-soft} shows a comparison in the HK- plane, along with line cuts along the H direction, between \LBCO\ and \NdLSCO. Immediately apparent is the far poorer correlation length in \NdLSCO, where we find that  the correlation length is again isotropic but in this case $\xi_H = 111\pm5$~\AA.

In both cases, the wavevector in the H direction, is 0.24~r.l.u. and is therefore incommensurate. These measurements were performed using an orientation matrix defined \emph{in-situ} by the $(002)_\mathrm{HTT}$ and $(101)_\mathrm{HTT}$ Bragg reflections, and as such their values are reported with high confidence. Previous studies have found similar incommensurabilities at these dopings (see e.g. Refs.~\onlinecite{Abbamonte:2005bl,Hucker:2011je,PhysRevB.54.7489}). We return to this point later.

The result that the charge stripes display isotropic correlation lengths is perhaps surprising given that the stripes in a single CuO$_2$ layer are highly anisotropic. While four-fold rotational symmetry is recovered when averaging over the whole crystal, since successive sheets along the c-axis are rotated by 90$^{\circ}$ with respect to each other, these present experiments only observe charge stripes with a propagation wavevector along one direction in reciprocal space - by virtue of the diffraction condition. There is therefore no {\em a priori} reason to expect these correlations to be isotropic. One possible explanation is that nearest-neighbor c-axis correlations couple the charge stripe correlations on alternating layers. It has been suggested, for example, that such inter-layer coupling could occur between the 1D metallic behavior of a stripe and the $\vec{q}$-vector in the neighboring stripe\cite{PhysRevB.56.6120}. Unfortunately, from our diffraction measurements it is not possible to say anything about nearest-neighbor c-axis coupling due to the 90$^\circ$ rotation of the stripes. A second possibility is that the isotropy is driven by poorly screened (long range) Coulomb interactions, as has been suggested in the context of a ladder compound\cite{Rusydi:2006bf}. Finally, and most prosaically,  it is possible that the charge stripe correlations are limited by intrinsic crystallographic disorder, for example due to defects arising from randomly distributed dopant ions\cite{PhysRevB.54.R15622}. As such order is random, one would expect it to be isotropic, and that it would therefore cause an isotropic correlation length. We will return to the role of crystallographic order again shortly in the discussion of modulation amplitudes.

We are also able to compare the peak intensities in the two systems. To do so, the intensities shown in Fig.~\ref{fig:comp-soft} have been normalized to the integrated intensity of the respective $(002)_\mathrm{HTT}$ Bragg reflections (measured at 1~keV) and corrected for their different structure factors. The integrated intensity, $V$, of a superlattice reflection is proportional to the product of the peak intensity and the widths in three orthogonal directions, $V \propto I_\mathrm{Peak}\times\Gamma_H\times\Gamma_K\times\Gamma_L$. As discussed, $\Gamma_K=\Gamma_H$, and we make the  assumption that the correlation lengths along L are equal in the two materials. The ratio of integrated intensities is then $V^{hole}_\mathrm{AB} = (I_{\mathrm{Peak}\;A} \times \Gamma^2_A) / (I_{\mathrm{Peak}\;B} \times \Gamma^2_B)$. Calculating from the data presented in Table~\ref{tab:results} we find that the ratio of integrated intensities, corrected for all experimental factors is $V^{hole}_\mathrm{rel} = 97\pm10$, i.e. the electronic charge stripe modulation in \LBCO\ is $\sim9.8\pm0.5$ times larger in amplitude than in \NdLSCO.

This factor of almost 10 is a surprisingly large number. Taken at face value it implies a very large difference in the value of the charge order parameter in these two systems and appears to contradict expectations based on measurements of the ordered magnetic moments in the two systems. $\mu$SR measurements, for example, show similar ordered moments of 0.3 $\mu_B$  with similarly large magnetically-ordered volume fractions in \LBCO\ and \NdLSCO\cite{PhysRevB.58.8760}. Since the magnetic order is constrained by the charge order, it would seem counterintuitive for the charge order parameters to be so different when the magnetic order parameters are so similar. Further, a recent estimate of the hole concentration per Cu site in the charge carrier stripes in La$_{1.675}$Eu$_{0.2}$Sr$_{0.125}$CuO$_4$  (LESCO) resulted in a value considerably larger than 0.2 holes\cite{PhysRevB.79.100502}. Based on the similarity of LESCO and \NdLSCO, in terms of correlation length and temperature dependence, once might expect LESCO and \NdLSCO\ to have similar hole modulations. Such an expectation is then, at first glance, inconsistent with the present measurement, which suggests \LBCO\ is 10 times larger, which would be an unphysical result. However, the  LESCO estimate came from an analysis of the resonant lineshape, which reflects the amplitude of the modulation in the modulated regions and there are a number of reasons why a measurement based on the diffraction signal could lead to a different result. Firstly, the ordered volume of sample could be different for \LBCO\ and \NdLSCO, with \NdLSCO having an ordered volume up to 100 times smaller than that of \LBCO. This seems extremely unlikely. A second possibility is that these diffraction measurements do not fully capture all the intensity of the stripe correlations and that in \NdLSCO\ there is significant intensity outside the detected region in reciprocal space - i.e. in diffuse tails. Missing this intensity would result in a much reduced measured hole amplitude in \NdLSCO\ compared with \LBCO. This effect would not be reflected in resonant lineshape analysis. Such an explanation is perhaps consistent with the fact that the charge stripes are more poorly correlated in \NdLSCO\ as compared to \LBCO. At this point, it is not possible to know which, if any, of these explanations is the correct one.

\begin{table}
  \caption{\label{tab:results}Comparison of hole and lattice modulation wavevectors, correlation lengths and peak intensities for \LBCO\ and \NdLSCO\ as measured with soft and hard x-rays respectively.}
  \begin{ruledtabular}
    \begin{tabular}{lllll}
      & & Center & HWHM& Peak\\
      & &  [r.l.u.]   & [r.l.u.] $\times 10^{-3}$&Intensity \\
      LBCO & 529.3~eV & 0.241 & $2.36\pm 0.05$  & $7.82\pm0.01$ \\
      LNSCO &  529.3~eV & 0.238 & $5.42\pm0.25$ & $(1.54\pm0.005)\times 10^{-3}$\\
      LBCO & 11.3 keV & - & $3.37\pm0.9$ & $0.225\pm0.002$ \\
      LNSCO & 11.3 keV & - & $5.59\pm0.9$ & $(4.17\pm0.18)\times 10^{-3} $\\
    \end{tabular}
  \end{ruledtabular}
\end{table}

\begin{figure}
  \includegraphics[width=0.8\columnwidth]{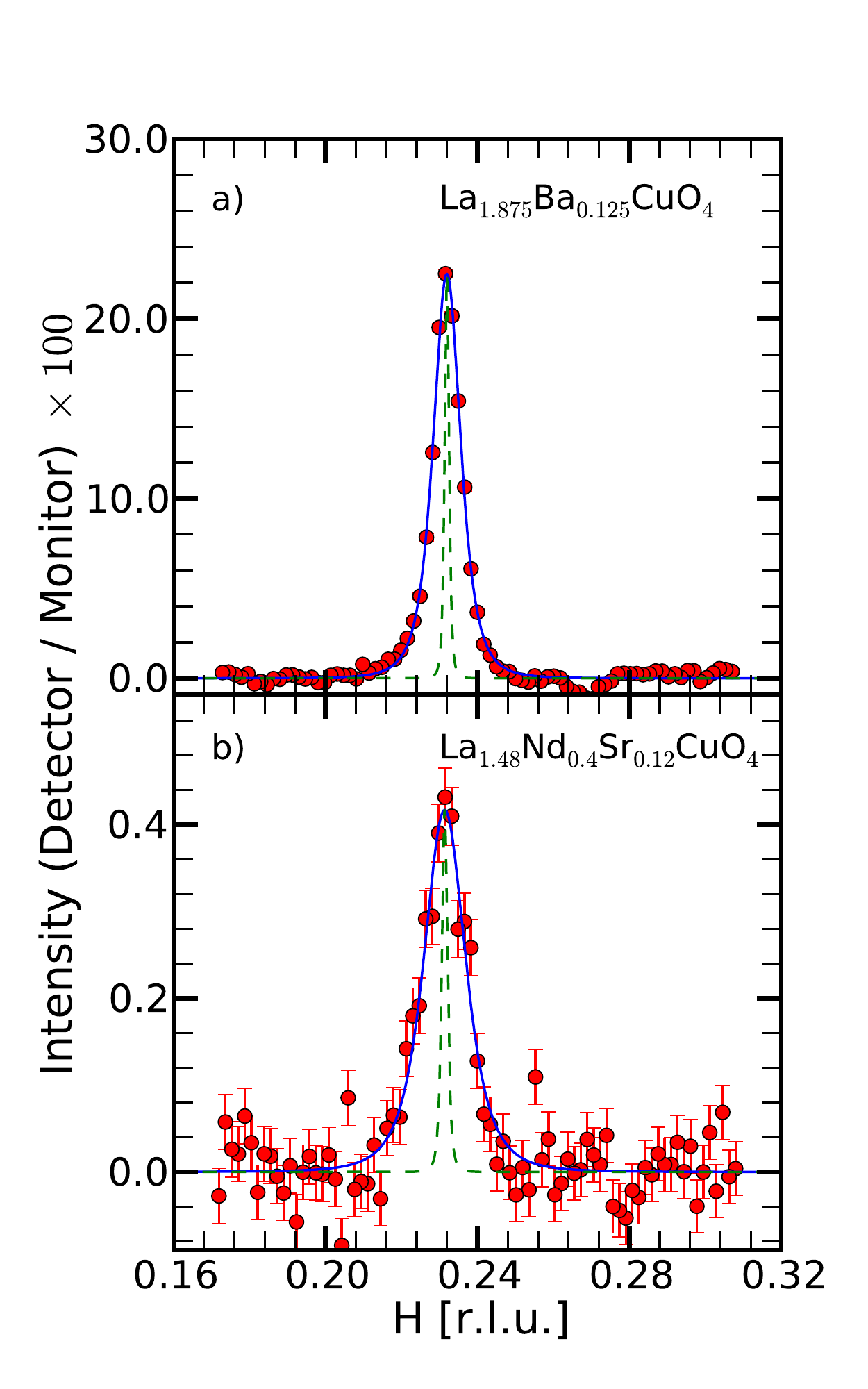}
  \caption{Comparison of stripe order peaks in a) \LBCO\ and b) \NdLSCO\
    as measured by hard x-rays, E$_i$=11.3~keV at T=12~K. Solid lines are the results of fits to Lorentzian squared lineshapes. Dashed lines represent the width of the $(1,0,12)_\mathrm{HTT}$ nearby Bragg reflection in each case, demonstrating the two samples are of similar quality. Scattered intensities are normalized to the $(1,0,12)_\mathrm{HTT}$ Bragg peak and therefore may be compared directly to each other.}
\label{fig:comp-hard}
\end{figure}

We next discuss the lattice modulation that is concomitant with the electronic modulation,  as measured with hard x-rays. In Fig.~\ref{fig:comp-hard}, we show scans performed with E$_i$=11.3~keV through the (0.24,0,11.5) charge stripe wavevector.  We have again normalized the scattered intensity by a nearby Bragg peak, in this case the $(1,0,12)_\mathrm{HTT}$. Due to the poorer experimental resolution of the hard x-ray measurements, it is necessary to deconvolve  the experimental resolution. This latter was determined from the $(1,0,12) _\mathrm{HTT}$ reflection. As an upper bound on the resolution, we took the sharpest of the \LBCO\ and \NdLSCO\ $(1,0,12) _\mathrm{HTT}$ reflections. After deconvolving the instrumental resolution we find $\xi^{lattice}_{LBCO}=178~\pm 50$~\AA\ and  $\xi^{lattice}_{LNSCO}=107~\pm 20$~\AA. Comparing the widths of the stripe peaks as measured with soft and hard x-rays, we find that the widths are equal within experimental error for the two x-ray energies in both \LBCO\ and \NdLSCO.

From the hard x-ray deconvolved width and the peak intensity, we can calculate
the ratio of the lattice modulations for the two systems following the
same arguments as before. We find that the ratio of the integrated
intensities of the lattice modulations is $V^{lattice}_\mathrm{rel} =
20\pm6$, i.e. the lattice modulation amplitude is $\sim4.4\pm0.7$
times larger in \LBCO\ than in \NdLSCO.

\begin{figure}
  \includegraphics[width=0.8\columnwidth]{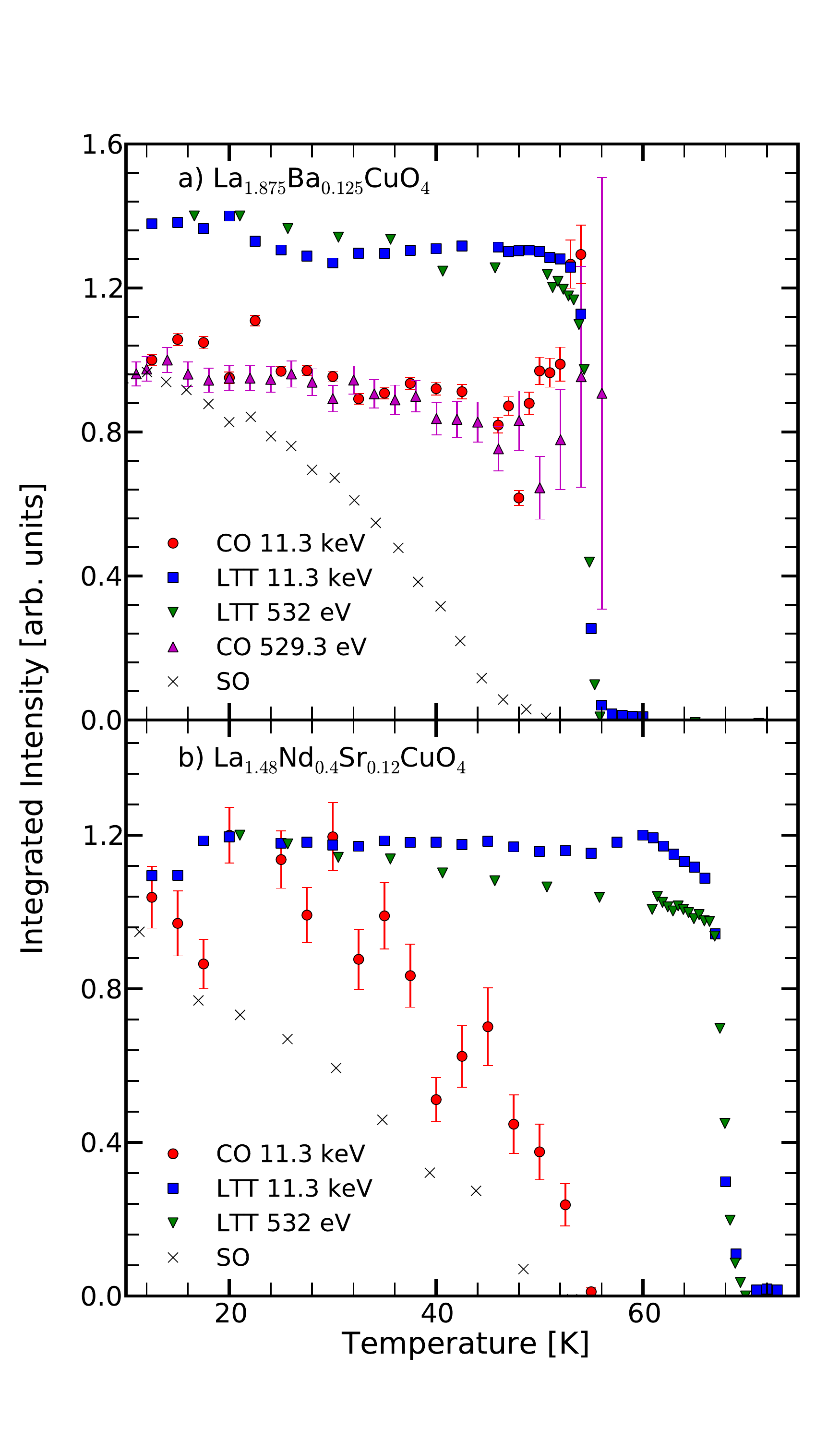}
  \caption{Temperature dependence of the lattice modulation (red circles), electronic modulation (magenta up triangles), spin modulation (black crosses) and the LTT phase as measured by soft (green down triangles) and hard x-rays (blue squares) in both \LBCO\ (a, top panel) and \NdLSCO\ (b, bottom panel). The data presented for the spin modulation were taken from Ref.~\onlinecite{Hucker:2011je} and Ref.~\onlinecite{PhysRevB.54.7489} for \LBCO\ and \NdLSCO, respectively. }
\label{fig:tdep}
\end{figure}

Before we turn to a discussion of the significance of these results, we present  the temperature dependence of the respective modulations, and of the LTT phase, for the two systems (Fig.~\ref{fig:tdep}). For soft x-rays, the latter is measured through the forbidden $(001)_\mathrm{HTT}$ reflection, observable only in the LTT phase at $E_i = 532$~eV\cite{PhysRevB.83.092503}.
For hard x-rays, the $(1,0,12)_\mathrm{HTT}$ reflection, observable only in the LTT phase is used.
In the case of the superlattice modulations, the integrated intensity
of the reflection is taken to be $\propto I\times\Gamma^2$. This is
consistent with two dimensional stripes, where the $L$-axis dependence
of the scattering is dominated by the structure factor due to the very
short correlation length\cite{vonZimmermann:1998tc}.

There are several important things to note from this plot. First, for
\LBCO\ it is clear that an observable peak is visible (both the strain
modulation and the charge density modulation) only within the LTT phase. In addition, the integrated intensity is
constant, within experimental error, for both the electronic and
structural modulations. This demonstrates that the amplitudes of both
these modulations are constant up to the LTT transition temperature, that is the order parameter is not changing, only the correlation length. This
suggests that had the LTT transition occurred at a higher temperature in \LBCO\, then
it is likely that both the electronic and structural modulations would
also have persisted to higher temperatures.

In contrast, for  \NdLSCO\ (Fig.~\ref{fig:tdep}b) the amplitude decays
upon warming and is unobservable above  $T_{\mathrm{CO}}$=55~K, well
below the LTT transition at $T_{\mathrm{LTT}}$=68~K.
This behavior is reminiscent of recent results in La$_{1.8}$Eu$_{0.2-x}$Sr$_{x}$CuO$_4$ which has an even larger difference between $T_{\mathrm{LTT}}$ and $T_{CO}$ of $\sim 55$~K\cite{PhysRevB.79.100502,PhysRevB.83.092503}.

Finally, it is worth noting that the temperature dependence of the
LTT-peaks are identical in both samples when measured at 11.3 ~keV and 530~eV,
demonstrating that the soft x-rays are probing the same bulk-like
behavior as the hard x-rays and that the samples are homogeneous.

The results presented above represent the first time these hole and
lattice modulations have been characterized in this detail, and as a result, we are
able to draw a number of conclusions. First, we address the incommensurability, that is the
deviation from the ideal value of $\delta = 0.25$ at this doping. In \NdLSCO\ this was previously attributed to a mixture of two different commensurate stripe spacings \cite{PhysRevB.59.14712}.  A random ordering of these components gave a good match to the spin and lattice incommensurabilities and correlation lengths determined with neutrons.  The considerably longer charge and lattice correlation lengths in \LBCO\ compared to \NdLSCO\, together with the same incommensurability in the two systems, is not compatible with this random-ordering model.  Within a model of two integral stripe spacings, correlated ordering of the two components would be necessary to explain these results.

We now address the differences between  \LBCO\ and \NdLSCO, searching
for hints as to the origin of the differences in stripe
amplitude. First, the average structures of \LBCO\ and \NdLSCO\ are
remarkably similar\cite{springerlink:10.1007/BF00754942,Katano:1993ef}.  Secondly, the electron-phonon coupling for the Cu-O bond stretching mode appears to be the same\cite{Reznik:2006iv}. Finally, the average A-site cation radius is again very similar: 1.232~\AA\ and 1.211~\AA\ for \LBCO\ and \NdLSCO\, respectively
\cite{Shannon:a12967}.  What then is the origin of the difference in the
two systems?  Local pinning of stripe domains (as occurs with Zn doping)\cite{PhysRevLett.91.067002} can cause neighboring stripe domains to be out of phase with one another, and destructive interference may limit correlation lengths and average amplitudes. The orthorhombic strains (associated with the octahedral tilts) at temperatures just above the LTT transition are twice as large in \NdLSCO\cite{PhysRevB.59.14712}  as in \LBCO\cite{Hucker:2011je} and are a candidate for the origin of this pinning. Indeed in the ordered phase the LTT tilt angle is larger in \NdLSCO\ than in \LBCO\ (4.4$^{\circ}$ compared to 3.5$^{\circ}$, see  Ref.~\onlinecite{springerlink:10.1007/BF00754942}) and it has previously been shown that the electronic properties are sensitive to this degree of tilt, with a critical value of $\Phi_c=3.6^{\circ}$ below which the electronic properties were unaffected\cite{PhysRevLett.73.1841}. By analogy with perovskite manganites\cite{PhysRevB.54.8992}, we may expect that the larger modulations of the Cu-O-Cu bond angles resulting from this tilt will be associated with a greater tendency to localize charge and perturb the stripe modulations, especially in the vicinity of divalent dopant ions.  Thus, the reduced average tilt angle in \LBCO\ might explain the longer correlation lengths and enhanced modulation amplitudes of the periodic stripes.

In summary, our comparison shows that the in-plane correlations of the charge stripes in these two systems are isotropic, and a factor of two larger in \LBCO. These two systems show different temperature dependences, with the integrated intensity of the charge modulation in \LBCO\ constant up to the LTT transition, while in \NdLSCO\ it decreases on warming, disappearing at $\sim 55$~K, well below the LTT transition in this material. Comparing the amplitudes of the lattice modulation we find that the lattice modulation in \LBCO\ is 4 times stronger than in \NdLSCO.  Performing the same comparison on the electronic modulation suggests that the electronic modulation, as measured by diffraction is 10 times stronger in \LBCO\ when compared with \NdLSCO. This result, based on measurement of the integrated intensity of the stripe modulation, appears at odds with measurements performed by resonant lineshape analysis. It is clear that further work is required to solve this puzzle, and we hope that this work encourages such measurements. 

We gratefully acknowledge D.S. Coburn, W. Leonhardt, W. Schoenig, and S. Wirick  for their
technical support. Work performed at BNL was supported by the US
Department of Energy, Division of Materials Science, under contract
No. DE-AC02-98CH10886.  J.G. appreciates the support by the DFG through the Emmy-Noether program.

\bibliographystyle{apsrev4-1}
\bibliography{../sbw_bib_short,../otherrefs}
\end{document}